\documentclass[a4paper,11pt]{article}
\pdfoutput=1 

\usepackage{jcappub} 
\usepackage{indentfirst}
\usepackage{appendix}
\usepackage{multirow}

\usepackage{array}
\usepackage{enumerate}
\usepackage{xcolor}
\usepackage{placeins}
\usepackage{graphicx}
\usepackage{dcolumn}
\usepackage{CJKutf8}
\usepackage{bm}
\usepackage{float}
\usepackage{amssymb}
\usepackage{latexsym}
\usepackage{booktabs}
\usepackage{amsmath}
\usepackage{subcaption}
\usepackage{url}
\begin{document}

\title{Assessing the consistency of the CPL dark-energy model with cosmological data: Simulation-calibrated neural posterior disagreement}

\author[a]{Tian-Yang Sun,}
\author[b]{Tian-Nuo Li,}
\author[a]{Guo-Hong Du,}
\author[c,d]{He Wang,}
\author[a]{Jing-Fei Zhang}
\author[a,e,f,1]{and Xin Zhang\note{Corresponding author.}}

\affiliation[a]{Liaoning Key Laboratory of Cosmology and Astrophysics, College of Sciences, Northeastern University, Shenyang 110819, China}
\affiliation[b]{School of Fundamental Physics and Mathematical Sciences, Hangzhou Institute for Advanced Study, UCAS, Hangzhou 310024, China}
\affiliation[c]{International Centre for Theoretical Physics Asia-Pacific, University of Chinese Academy of Sciences, Beijing 100190, China}
\affiliation[d]{Taiji Laboratory for Gravitational Wave Universe (Beijing/Hangzhou), University of Chinese Academy of Sciences, Beijing 100190, China}
\affiliation[e]{Key Laboratory of Data Analytics and Optimization for Smart Industry (Ministry of Education), Northeastern University, Shenyang 110819, China}
\affiliation[f]{National Frontiers Science Center for Industrial Intelligence and Systems Optimization, Northeastern University, Shenyang 110819, China}

\emailAdd{suntianyang@stumail.neu.edu.cn, litiannuo@ucas.ac.cn, duguohong@stumail.neu.edu.cn, hewang@ucas.ac.cn, jfzhang@mail.neu.edu.cn, zhangxin@mail.neu.edu.cn}

\abstract{
Recent DESI observations hint at a preference for dynamical dark energy with an equation of state that crosses the phantom divide within the Chevallier-Polarski-Linder (CPL) parameterization. In this work, we assess the consistency of the CPL parameterization with current cosmological observations. We develop a model check based on the disagreement between posteriors inferred by neural posterior estimation and flow-matching posterior estimation. Both estimators are trained on the same CPL simulations, and their posterior disagreement is calibrated using independent simulations to obtain a global $p$-value. Controlled simulations assess the reliability of the calibrated \(p\)-values and show how detection sensitivity depends on the form of departure from the assumed model. Applying the test to the compressed CMB+DESI+DESY5 data vector yields $p_{\rm global}=0.524$, indicating that the current observations are consistent with the CPL model. The method has a median evaluation time of only $0.53\,{\rm s}$, enabling efficient assessments of cosmological model adequacy. This framework enables efficient tests of dark-energy models, providing a firmer basis for investigating the physical origin of cosmic acceleration.
}

\maketitle

\section{Introduction}
\label{sec:introduction}

The \(\Lambda\) cold dark matter (\(\Lambda\)CDM) model explains late-time acceleration with a cosmological constant, \(w=-1\). It also successfully describes a broad range of cosmological observations, including the cosmic microwave background (CMB), baryon acoustic oscillations (BAO), and type Ia supernova (SN)~\cite{Planck:2018vyg,DESI:2024mwx,DES:2024jxu}. Nevertheless, the physical origin of cosmic acceleration and the reason for the small vacuum-energy scale remain unresolved~\cite{Weinberg:1988cp,Copeland:2006wr,Frieman:2008sn,Huterer:2017buf,Peebles:2002gy,Carroll:2000fy,Li:2011sd,Weinberg:2013agg,Martin:2012bt}. In addition, observational discrepancies, particularly the \(H_0\) tension~\cite{Planck:2018vyg,Riess:2019qba,Riess:2021jrx,Verde:2019ivm,Guo:2018ans,DiValentino:2019ffd,Vagnozzi:2019ezj,DiValentino:2020zio,DiValentino:2021izs,Vagnozzi:2021gjh,Abdalla:2022yfr,Schoneberg:2021qvd,Giare:2023xoc,Hu:2023jqc,Poulin:2023lkg,Vagnozzi:2023nrq,Cai:2026swf,Schoneberg:2026vaf,Sabogal:2026ipu,Pedrotti:2026dwj,Pedrotti:2025ccw,Liu:2025evk}, have motivated the exploration of physics beyond \(\Lambda\)CDM~\cite{Joyce:2014kja,Clifton:2011jh,Xu:2016grp,Gao:2022ahg,Gao:2021xnk,Zhang:2025dwu,Zhang:2005rj,Zhang:2005rg,Zhang:2006av,Boisseau:2000pr,Chevallier:2000qy,Li:2004rb,Zhang:2005hs,Zhang:2005yz,Zhang:2007sh,Ma:2007av,Zhang:2009un,Zhang:2014ifa,Li:2014eha,Cai:2015emx,Wang:2016lxa,Feng:2017usu,Poulin:2018cxd,Pan:2019gop,Pan:2020zza,Li:2023gtu}.

Recent BAO measurements from the Dark Energy Spectroscopic Instrument (DESI) have further challenged the $\Lambda$CDM model~\cite{DESI:2024mwx}. The DESI Data Release 2 (DR2) BAO measurements combined with CMB data and the SN sample show a $2.8$-$4.2\sigma$ preference for dynamical dark energy within the Chevallier-Polarski-Linder (CPL) parameterization~\cite{DESI:2025zgx,DESI:2025fii}. These DESI results have sparked an extensive discussion of the nature of dark energy and the robustness of evidence for its dynamical evolution~\cite{Jiang:2024xnu,Giare:2024smz,Giare:2024gpk,Lemos:2025bgy,Pang:2025lvh,Ye:2024ywg,Liu:2025mub,Li:2025ops,You:2025uon,Li:2026xaz,Li:2026hwq,Du:2026qtq,Li:2025owk,Turner:2026vzp,Ribeiro:2026lyd,Bella:2026zuk,Wang:2024dka,Giare:2026oti,Wang:2026wrk,Wang:2025znm,Yang:2025gaz,Cai:2025mas,Song:2025bio,Ren:2026jyw,Wang:2026vqw,Yang:2026qmz,Pan:2025qwy,Yang:2026yaq,Du:2025xes,Pedrotti:2026ncd,Yao:2025twv,Mukherjee:2026lnb,Yang:2026vti,Zhang:2026axl,Zheng:2026tye,Babu:2026tfr,Wu:2025vrl,Wu:2026qog,Wu:2025wyk}. For a recent review, see Ref.~\cite{Li:2026asg}.

The CPL parameterization is a widely used two-parameter description of evolving dark energy~\cite{Chevallier:2000qy,Linder:2002et}. Although it improves the fit over $\Lambda$CDM for several current data combinations~\cite{DESI:2025fii}, its linear dependence on the scale factor limits the dark-energy histories it can describe. Rapid changes in the equation of state, for example, may not be accurately reconstructed by CPL~\cite{Linden:2008mf,Ma:2011nc,Li:2011dr,Nesseris:2025lke}. Moreover, different parameterizations and physical models can produce similar background evolution~\cite{Feng:2004ad,Jassal:2005qc,Zhang:2005kj,Barboza:2008rh,Fu:2011ab,Pan:2019brc,Cheng:2025lod,Paul:2025wix,Li:2024qso,Li:2024qus,Li:2025eqh,Li:2025dwz,Li:2025ula,Li:2025owk,Hannestad:2002ur,Zlatev:1998tr,Zhang:2006qu,Li:2026xaz,Scherrer:2015tra,Li:2025vuh}. Therefore, it is necessary to further test the validity of these dark-energy models.

Commonly, model comparison, goodness-of-fit statistics, and posterior predictive checks are used to assess cosmological models~\cite{vonWietersheim-Kramsta:2024cks,Li:2010xjz,Trotta:2008qt,DES:2020lei,Romero-Shaw:2022ctb}. However, model comparison assesses only the support for candidate models relative to the \(\Lambda\)CDM model and cannot, on the basis of the comparison results alone, establish whether a model adequately describes the observations. In addition, checks that require repeated posterior inference can be computationally demanding, especially when the forward model is costly.

To address this computational challenge, amortized simulation-based inference (SBI) moves much of this cost to an offline training stage, allowing a trained estimator to be reused for new simulations and observations~\cite{Talts:2018zdk,Hermans:2021rqv,Dax:2021tsq,Dax:2022pxd,Sun:2023vlq,Dax:2024mcn,DBLP:journals/corr/abs-2607-21702,DBLP:conf/nips/PapamakariosM16,DBLP:conf/icml/GreenbergNM19,DBLP:conf/icml/HermansBL20,Cranmer:2019eaq,DBLP:journals/jmlr/PapamakariosNRM21,DBLP:conf/aistats/LueckmannBGGM21,DBLP:conf/nips/WildbergerDBGMS23,Sun:2025ypd,Qin:2025mvj,Kofler:2025dux,Sun:2026gsq}. This amortization has enabled likelihood-free cosmological inference with expensive simulators~\cite{Alsing:2019xrx,Cole:2021gwr,Hahn:2023udg,Surrao:2024lgo,Sun:2024ywb,Pierre:2025ulp,Alsing:2018eau,Brehmer:2018hga,Jeffrey:2020xve,Wang:2022qta,Wang:2023vej,DES:2024xij}. It also makes it feasible to compare the posteriors returned by different estimators across large ensembles of data.

Such posterior comparisons can inform model checking, as different inference methods can respond differently when the assumed model does not adequately describe the data~\cite{DBLP:journals/corr/abs-2209-01845,DBLP:journals/tmlr/ElsemullerOSBKR24,DBLP:journals/corr/abs-2406-03154,DBLP:conf/icml/WehenkelGSBSJC25,DBLP:journals/corr/abs-2503-12315,Pierre:2025ulp,Liu:2026nhu,Sun:2026dga}. Previous work has used Kullback-Leibler (KL) divergences between independently trained neural posterior estimators to identify possible model misspecification~\cite{Alvey:2025hdv}. However, a KL divergence alone does not indicate how unusual the observed disagreement is under the assumed model. Other work has developed statistically calibrated model checks using neural classifiers to identify specific discrepancies between observed data and simulations~\cite{AnauMontel:2024flo}. However, these approaches do not directly provide a calibrated statistical test of disagreement between different posterior estimators.

In this work, we check the consistency of the CPL dark-energy model with cosmological observations. Our simulation-calibrated test is based on the disagreement between posteriors inferred by neural posterior estimation (NPE) and flow matching posterior estimation (FMPE), both trained on the same CPL simulations. A first ensemble of additional CPL simulations establishes reference distributions for three complementary posterior-distance measures, yielding individual $p$-values whose minimum defines a combined statistic. Interpreting this minimum as a single-test $p$-value can overstate the significance of the disagreement. We therefore use a second, independent CPL ensemble to establish the reference distribution of the combined statistic and obtain a global $p$-value that accounts for the selection across measures. Separate validation simulations assess the calibration and detection sensitivity using controlled examples and alternative dark-energy histories. We then apply the test to a compressed data vector combining Planck CMB, DESI DR2 BAO, and DESY5 SN observations.

This work is organized as follows. Section~\ref{sec:method} describes the CPL data model, neural posterior estimators, and posterior-disagreement test. Section~\ref{sec:cosmo_results} presents the illustrative, simulated, and observed-data results. Section~\ref{sec:conclusion} summarizes the conclusions. The appendices describe the construction of the compressed cosmological data vector, finite-sample calibration, posterior-coverage diagnostics, and the implementation and supplementary results of the controlled-validation experiments.

\section{Method}
\label{sec:method}

\subsection{Chevallier-Polarski-Linder parameterization}
\label{sec:cpl_cosmology}

We test the spatially flat CPL parameterization~\cite{Chevallier:2000qy,Linder:2002et},
\begin{equation}
 w(a)=w_0+w_a(1-a),
 \label{eq:cpl}
\end{equation}
where $a=(1+z)^{-1}$. The parameters $w_0$ and $w_a$ specify the present value and scale-factor dependence of the dark-energy equation of state, respectively.

In addition to $w_0$ and $w_a$, we vary the physical baryon and cold-dark-matter densities $\omega_b$ and $\omega_c$, the Hubble constant $H_0$, and the reionization optical depth $\tau$. The primordial curvature power spectrum is described by its amplitude $A_s$ and spectral index $n_s$. We define $\omega_b=\Omega_bh^2$, $\omega_c=\Omega_ch^2$, and $h=H_0/(100\,{\rm km\,s^{-1}\,Mpc^{-1}})$. The total matter density parameter $\Omega_m=\Omega_b+\Omega_c+\Omega_\nu$ is derived, with $\Omega_\nu$ denoting the massive-neutrino contribution. The full parameter vector is
\begin{equation}
 \bm{\theta}_{\rm CPL}=
 (\omega_b,\omega_c,H_0,\tau,\ln(10^{10}A_s),n_s,w_0,w_a).
 \label{eq:cpl_parameters}
\end{equation}

We evolve CPL perturbations with the parameterized post-Friedmann (PPF) prescription in \texttt{CAMB}~\cite{Lewis:1999bs,Fang:2008sn}. The parameter conventions follow the \texttt{Cobaya} baseline configuration~\cite{Torrado:2020dgo}. The calculation assumes zero spatial curvature and adiabatic scalar initial conditions. It also assumes \(N_{\rm eff}=3.044\), one massive neutrino with total mass \(0.06\) eV, and dark-energy sound speed \(c_s^2=1\). The helium abundance follows the \texttt{CAMB} big-bang-nucleosynthesis prescription. Table~\ref{tab:cosmo_priors} lists the simulation-prior bounds. We retain only draws satisfying $w_0+w_a\leq0$ and yielding finite PPF solutions in \texttt{CAMB}.

\begin{table}[t]
  \centering
  \small
  \caption{The prior bounds used to simulate the data of the CPL model.}
  \label{tab:cosmo_priors}
  \begin{tabular}{lcc}
    \toprule
    Parameter & Lower & Upper \\
    \midrule
    \(\omega_b\) & 0.019 & 0.026 \\
    \(\omega_c\) & 0.08 & 0.16 \\
    \(H_0\,[{\rm km\,s^{-1}\,Mpc^{-1}}]\) & 50 & 90 \\
    \(\tau\) & 0.01 & 0.12 \\
    \(\ln(10^{10}A_s)\) & 2.7 & 3.6 \\
    \(n_s\) & 0.90 & 1.05 \\
    \(w_0\) & -3.0 & 0.5 \\
    \(w_a\) & -5.0 & 3.0 \\
    \bottomrule
  \end{tabular}
\end{table}

\subsection{Cosmological data and simulations}
\label{sec:data_simulation}

We combine CMB temperature and polarization spectra, DESI DR2 BAO distances, and the DESY5 SN observation. The original CMB spectra and SN observation are high-dimensional. To reduce the neural input dimension and the cost of generating and processing millions of simulated data vectors, we apply a fixed compression scheme throughout the analysis.

For the CMB data, we use the Planck 2018 temperature auto-spectrum (TT), temperature-E-mode polarization cross-spectrum (TE), and E-mode polarization auto-spectrum (EE) over $30\leq\ell\leq900$~\cite{Planck:2019nip}. We generate mock spectra following Refs.~\cite{Wang:2025xjc,Wang:2026ske}, with the joint temperature-polarization covariance described in Ref.~\cite{Percival:2006ss}. The fixed covariance incorporates the Planck best-fit theory spectrum~\cite{Planck:2018vyg}, cosmic variance, and the instrumental-noise model implemented in \texttt{MontePython}~\cite{Audren:2012wb,Brinckmann:2018cvx}. We compress each spectrum into 18 multipole bins, yielding a 54-dimensional CMB data vector. The same binning is applied to the observed and simulated spectra, with the covariance propagated accordingly.

The BAO data comprise the 13 observables and released covariance of the DESI DR2 sample~\cite{DESI:2025zgx}. Each coordinate measures \(D_M/r_d\), \(D_H/r_d\), or \(D_V/r_d\), depending on tracer and redshift. Here, \(D_M\) is the transverse comoving distance and \(D_H=c/H(z)\) is the Hubble distance. The quantity \(r_d\) is the sound horizon at the drag epoch.

The SN data use the original DESY5 release~\cite{DES:2024jxu}, including the redshifts, distance moduli, statistical uncertainties, and systematic covariance. To reduce the neural-network input dimension, we compress the distance-modulus vector into 24 redshift bins using inverse statistical-variance weights. The full statistical-plus-systematic covariance is propagated through the same linear transformation. We then project out the common magnitude offset, yielding a 23-dimensional SN data vector that is insensitive to the absolute-magnitude zero point. The same compression and projection are applied to the observed and simulated data.

Additional details of the CMB conversion and linear SN compression are given in Appendix~\ref{app:implementation}. The resulting Planck+DESI+DESY5 data vector, hereafter referred to as CBS, has 90 components, comprising 54 CMB band powers, 13 BAO observables, and 23 compressed SN components. Every cosmological check uses the full vector. We generate mock data vectors by adding Gaussian noise to the CPL model predictions,
\begin{equation}
\bm{x}=\bm{\mu}_{\rm CPL}(\bm{\theta})+\bm{\epsilon},
\qquad \bm{\epsilon}\sim{\cal N}(\bm{0},C),
\label{eq:cosmo_mock}
\end{equation}
where $\bm{\mu}_{\rm CPL}(\bm{\theta})$ is the theoretical CBS data vector for cosmological parameters $\bm{\theta}$, and $C$ contains the fixed CMB, BAO, and SN covariance data described above. The same fixed covariance is used to generate the simulated noise and to analyze the observed data vector.

\subsection{Simulation-based inference}
\label{sec:sbi}

The posterior distribution describes the uncertainty in model parameters after accounting for the data. In SBI, neural networks learn this distribution from simulated parameter-data pairs and can then analyze new data without running a new Markov chain Monte Carlo (MCMC) analysis. We train NPE and FMPE on the same simulations, so both estimators target the same conditional posterior but optimize different objectives. Their posterior differences provide the discrepancy statistic calibrated in section~\ref{sec:disagreement_test}.

NPE directly approximates the conditional posterior~\cite{DBLP:conf/icml/GreenbergNM19}. We denote this approximation by \(q_\phi(\bm{\theta}|\bm{x})\). For simulations drawn from \(\pi(\bm{\theta})p(\bm{x}|\bm{\theta})\), NPE minimizes
\begin{equation}
 {\cal L}_{\rm NPE}(\phi)
 =-\mathbb{E}_{\pi(\bm{\theta})p(\bm{x}|\bm{\theta})}
 \!\left[\log q_\phi(\bm{\theta}|\bm{x})\right].
 \label{eq:npe_loss}
\end{equation}
The trained density can be sampled directly and evaluated pointwise.

FMPE learns a conditional vector field that transports a simple base density to the posterior~\cite{DBLP:conf/nips/WildbergerDBGMS23}. Its loss compares the learned field \(v_\psi\) with the target field \(u_t\) along a prescribed probability path,
\begin{equation}
 {\cal L}_{\rm FMPE}(\psi)
 =\mathbb{E}\!\left[
 \left\|v_\psi(t,\bm{z}_t,\bm{x})
 -u_t(\bm{z}_t|\bm{\theta})\right\|_2^2\right].
 \label{eq:fmpe_loss}
\end{equation}
Here $\bm{z}_t$ denotes a point on the probability path at time $t$. We implement NPE and FMPE using the \texttt{sbi} package~\cite{BoeltsDeistler_sbi_2025}. The NPE density is represented by a masked autoregressive flow~\cite{Papamakarios:2017tec}.

We also use neural ratio estimation (NRE) in the low-dimensional validation examples. NRE formulates likelihood-free inference as a binary classification problem between samples from the joint distribution $p(\bm{\theta},\bm{x})$ and independently paired samples from $p(\bm{\theta})p(\bm{x})$. Let $d_{\eta}(\bm{\theta},\bm{x})$ denote the classifier probability that a pair belongs to the joint-distribution class. For equal class probabilities, the classifier odds estimate the likelihood-to-evidence ratio,
\begin{equation}
\frac{d_{\eta}(\bm{\theta},\bm{x})}
{1-d_{\eta}(\bm{\theta},\bm{x})}
\simeq
\frac{p(\bm{x}\mid\bm{\theta})}{p(\bm{x})},
\end{equation}
from which the posterior follows as $p(\bm{\theta}\mid\bm{x})\propto\pi(\bm{\theta})d_{\eta}/(1-d_{\eta})$~\cite{DBLP:conf/icml/HermansBL20,Cranmer:2019eaq}. Marginal and truncated variants of NRE have also been applied to cosmological inference~\cite{Cole:2021gwr}. In our low-dimensional examples, we evaluate the learned ratio on a dense parameter grid and obtain posterior samples by grid resampling. For the cosmological analysis, NRE would require an additional posterior-sampling procedure, for each simulated data vector, making inference over the large mock ensembles needed for calibration computationally expensive. We therefore use NPE and FMPE, which allow direct sampling from the full posterior over cosmological parameters.

The cosmological analysis uses one NPE-FMPE pair trained on the compressed CBS simulations. Table~\ref{tab:sbi_settings} summarizes the architectures and optimization settings, and ReLU denotes the rectified linear unit activation.

\begin{table}[H]
  \centering
  \small
  \caption{Training settings for the CPL posterior estimators used in the compressed CBS analysis.}
  \label{tab:sbi_settings}
  \begin{tabular}{lcc}
    \toprule
    Setting & NPE & FMPE \\
    \midrule
    Hidden layers & 6 & 6 \\
    Hidden width & 256 & 256 \\
    Autoregressive transforms & 8 & - \\
    Activation & ReLU & ReLU \\
    Optimizer & AdamW & AdamW \\
    Batch size & 1024 & 1024 \\
    Initial learning rate & \(5\times10^{-4}\) & \(5\times10^{-4}\) \\
    Training epochs & 100 & 100 \\
    \bottomrule
  \end{tabular}
\end{table}

For each epoch, 60,000 new data samples with Gaussian noise are generated. Both NPE and FMPE networks use cosine learning-rate decay to \(10^{-5}\). NPE and FMPE share the theory banks and noisy training batches. Their network parameters, objectives, and optimizers remain separate. We standardize parameters during training and use min-max normalized coordinates for posterior distances. A bijective transformation maps the bounded CPL parameter domain to unconstrained network coordinates and enforces \(w_0+w_a\leq0\). The posterior samples therefore satisfy the bounds in Table~\ref{tab:cosmo_priors} and this inequality.

\subsection{Posterior-disagreement test}
\label{sec:disagreement_test}

For each data vector, we compare the posteriors inferred by NPE and FMPE. A large posterior distance is considered anomalous only relative to the distances obtained from data generated under the CPL model. We establish this reference using two independent calibration ensembles, ${\cal R}$ and ${\cal O}$, separate from the training data. The first ensemble, ${\cal R}$, converts each posterior distance into an individual $p$-value. The second, ${\cal O}$, calibrates the minimum of these values to account for the selection across distance measures. Applying the same procedure to the observed data yields the global $p$-value $p_{\rm global}$.

\begin{figure}[H]
  \centering
  \includegraphics[width=\textwidth]{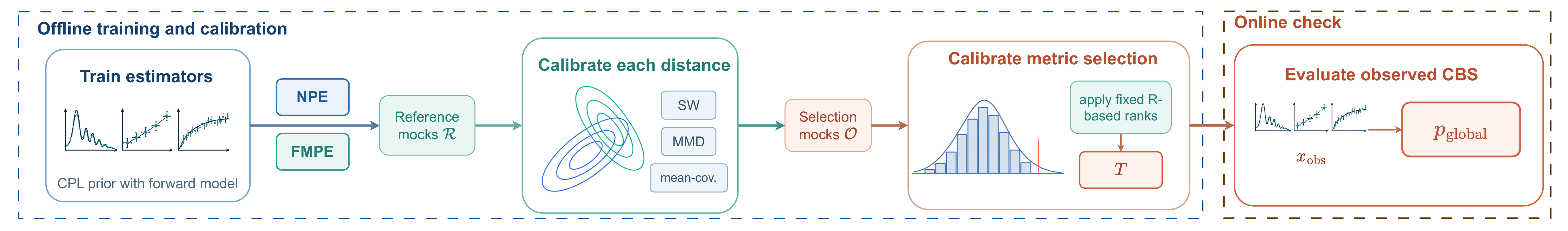}
  \caption{Workflow of the posterior-disagreement model check. Offline CPL simulations train NPE and FMPE. Reference mocks ${\cal R}$ convert each posterior distance into an individual $p$-value, and independent mocks ${\cal O}$ calibrate the minimum across distance measures. The observed CBS vector follows the same fixed procedure to obtain $p_{\rm global}$.}
  \label{fig:method_workflow}
\end{figure}

Let \(\pi_{\rm CPL}(\bm{\theta})\) be the simulation prior and \(p_{\rm CPL}(\bm{x}|\bm{\theta})\) the data model. The prior-predictive distribution is
\begin{equation}
  p_{\rm CPL}(\bm{x})=
  \int p_{\rm CPL}(\bm{x}|\bm{\theta})
  \pi_{\rm CPL}(\bm{\theta})\,d\bm{\theta}.
  \label{eq:prior_predictive}
\end{equation}
We use the CPL prior-predictive distribution as the reference, averaging over the simulation prior rather than fixing the cosmological parameters. We assess consistency with this reference through the calibrated posterior disagreement. Figure~\ref{fig:method_workflow} summarizes the calculation.

Let \(q_{\rm N}(\bm{\theta}|\bm{x})\) and \(q_{\rm F}(\bm{\theta}|\bm{x})\) denote the NPE and FMPE posteriors. We use three complementary distances~\cite{DBLP:journals/jmlr/GrettonBRSS12,DBLP:conf/nips/KolouriNSBR19}. The sliced-Wasserstein distance compares one-dimensional projections of the posteriors, the maximum mean discrepancy (MMD) with a radial basis function (RBF) kernel compares kernel-averaged similarities within and between posterior samples, and the mean-covariance distance isolates differences in the first two moments. For samples \(\{\bm{\theta}_i\}_{i=1}^{n}\sim q_{\rm N}\) and \(\{\bm{\vartheta}_j\}_{j=1}^{n}\sim q_{\rm F}\), the sliced-Wasserstein (SW) distance is
\begin{equation}
  D_{\rm SW}(q_{\rm N},q_{\rm F})=
  \frac{1}{L}\sum_{\ell=1}^{L}
  W_1\!\left(
  \{\bm{u}_\ell^{\mathsf T}\bm{\theta}_i\}_{i=1}^n,
  \{\bm{u}_\ell^{\mathsf T}\bm{\vartheta}_j\}_{j=1}^n
  \right),
\end{equation}
where $W_1$ is the one-dimensional Wasserstein distance, and $\bm{u}_{\ell}$ are $L=256$ random unit vectors. The corresponding RBF-MMD is
\begin{align}
  D_{\rm MMD}^2(q_{\rm N},q_{\rm F})
  &=\frac{1}{n^2}\sum_{i,j} k(\bm{\theta}_i,\bm{\theta}_j)
  +\frac{1}{n^2}\sum_{i,j} k(\bm{\vartheta}_i,\bm{\vartheta}_j)
  \nonumber\\
  &\quad-\frac{2}{n^2}\sum_{i,j} k(\bm{\theta}_i,\bm{\vartheta}_j),
  \qquad
  k(\bm{y},\bm{z})=\exp[-\gamma\|\bm{y}-\bm{z}\|^2].
  \label{eq:mmd_distance}
\end{align}
The mean-covariance distance isolates the first two moments,
\begin{equation}
  D_{\mu C}(q_{\rm N},q_{\rm F})=
  \|\hat\mu_{\rm N}-\hat\mu_{\rm F}\|_2
  +\frac{1}{2}\|\hat C_{\rm N}-\hat C_{\rm F}\|_F.
\end{equation}
For the cosmological test, all parameters are mapped to dimensionless min-max coordinates before the distances are evaluated. The 256 SW directions are drawn once and reused for every mock and observation throughout the cosmological analysis. For the RBF kernel, \(\gamma^{-1}\) is the median nonzero squared separation among 2048 pre-specified prior samples. This bandwidth is also fixed throughout the cosmological analysis. Equation~\eqref{eq:mmd_distance} is the biased estimator of \({\rm MMD}^2\), evaluated with at most 1024 draws from each posterior.

For a data vector $\bm{x}$, the posterior distance associated with measure $m$ is
\begin{equation}
  S_m(\bm{x})=
  D_m\!\left[q_{\rm N}(\bm{\theta}|\bm{x}),
  q_{\rm F}(\bm{\theta}|\bm{x})\right].
\end{equation}

The reference ensemble ${\cal R}$ converts each posterior distance into an individual $p$-value through its empirical upper-tail rank~\cite{AnauMontel:2024flo},
\begin{equation}
  p_m(\bm{x};{\cal R})=
  \frac{1+\sum_{\bm{x}_r\in{\cal R}}
  \mathbf{1}[S_m(\bm{x}_r)\geq S_m(\bm{x})]}
  {|{\cal R}|+1}.
  \label{eq:ps}
\end{equation}
A small $p_m$ means that the estimators disagree more than they do for most CPL mocks. We define the combined statistic from the smallest individual $p$-value,
\begin{equation}
  T(\bm{x})=-\log_{10}\!\left[\min_m p_m(\bm{x};{\cal R})\right].
\end{equation}
Because the minimum is selected from three individual $p$-values, it cannot be interpreted directly as the global $p$-value. We therefore calibrate the combined statistic $T$ using a second, independent ensemble ${\cal O}$,
\begin{equation}
  p_{\rm global}=
  \frac{1+\sum_{\bm{x}_j\in{\cal O}}
  \mathbf{1}[T(\bm{x}_j)\geq T(\bm{x}_{\rm obs})]}
  {|{\cal O}|+1}.
  \label{eq:global_p}
\end{equation}
This two-stage calibration accounts for the selection across multiple posterior-distance measures~\cite{AnauMontel:2024flo}. The observed and simulated data are processed with the same fixed inference and distance-evaluation rules. Under the CPL reference model, the resulting Monte Carlo $p$-value is valid for finite calibration ensembles. The calibration argument and treatment of tied values are given in Appendix~\ref{app:finite_mock}.

For the cosmological calculation, each estimator contributes 1024 posterior samples, with the same draw count and score-generating rule applied to the observation and calibration mocks. For the cosmological analysis, we use 2000 simulations in \({\cal R}\), 800 in \({\cal O}\), and 400 held-out null mocks.

We flag unusually large posterior disagreement when $p_{\rm global}\leq0.05$. We benchmark the online calculation after loading the trained estimators and calibration results. Each timing includes posterior sampling, evaluation of the three distances, and empirical rank lookup, using the same eight-parameter CPL estimators and posterior sample count as the observed-data analysis. We perform 20 repeats on an NVIDIA RTX A6000, excluding the offline costs of simulation, training, and calibration.

The posterior diagnostics and the model check assess different properties. Highest-posterior-density (HPD) ranks on independent simulations assess model-averaged posterior coverage, whereas $p_{\rm global}$ measures whether the disagreement between the two estimated posteriors is unusually large under the CPL reference distribution. Appendix~\ref{app:posterior_diagnostics} gives the corresponding results.

\section{Results and discussion}
\label{sec:cosmo_results}

\subsection{Controlled validation of the posterior-disagreement test}
\label{sec:illustrative_examples}

We first compare two controlled departures from the training model: a one-dimensional affine displacement \(g_\delta(\theta)=\theta+\delta\) and a two-dimensional nonlinear deformation \(k_\delta(\bm{\theta})=\bm{f}(\bm{\theta})+\delta\bm{w}(\bm{\theta})\). In both cases, \(\delta=0\) gives the training model, and increasing \(\delta\) increases the departure. The simulator definitions and numerical settings are given in Appendix~\ref{app:illustrative_details}.

NPE, FMPE, and NRE are trained on the corresponding unperturbed simulator. Combining their three estimator pairs with the three posterior distances gives nine scores, whose minimum individual \(p\)-value is calibrated using independent simulations as in section~\ref{sec:disagreement_test}.

At the nominal rejection level of 0.05, the calibrated global test rejects 24 of 500 null observations in the affine example and 15 of 500 in the nonlinear example, giving rejection fractions of 0.048 and 0.030, respectively. For the nine individual estimator-pair and distance scores, the mean marginal rejection fractions are 0.0533 and 0.0537. The marginal-rank distributions are shown in Appendix~\ref{app:illustrative_details}.

Figure~\ref{fig:toy_geometry} shows how the estimated posteriors respond at $\delta=1.5$. In the affine example, all three estimators produce similar posterior shifts. Since the three distance measures are invariant under an exactly shared translation, a common parameter shift need not increase posterior disagreement. In the nonlinear example, the posterior locations, widths, and orientations change by different amounts across estimators, producing greater disagreement. The diagnostic therefore responds to differences between estimator responses, rather than to posterior displacement alone. The support representations used in these examples are described in Appendix~\ref{app:illustrative_details}.

\begin{figure}[H]
  \centering
  \includegraphics[width=\textwidth]{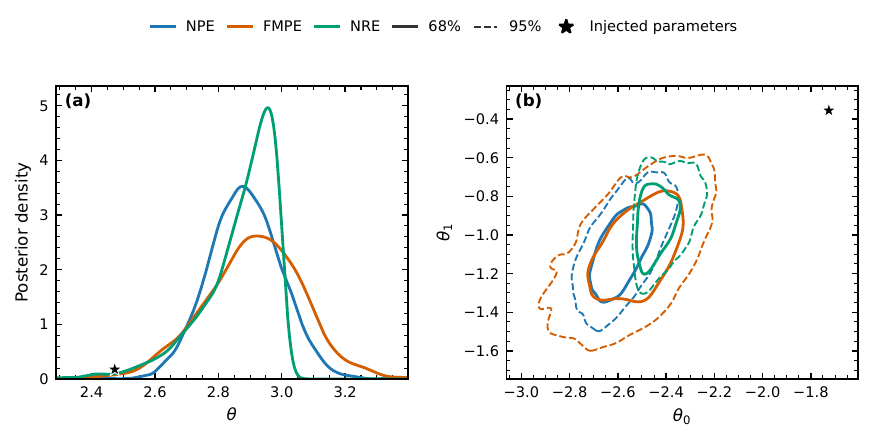}
  \caption{Posterior behaviour at $\delta=1.5$. (a) One-dimensional posterior densities for the affine displacement $g_{1.5}$. (b) Two-dimensional posterior contours for the nonlinear deformation $k_{1.5}$. Colours identify NPE, FMPE, and NRE. Solid and dashed contours in panel (b) enclose 68\% and 95\% of the posterior mass, respectively, and stars mark the injected parameters.}
  \label{fig:toy_geometry}
\end{figure}

\begin{figure}[H]
  \centering
  \includegraphics[width=0.94\textwidth]{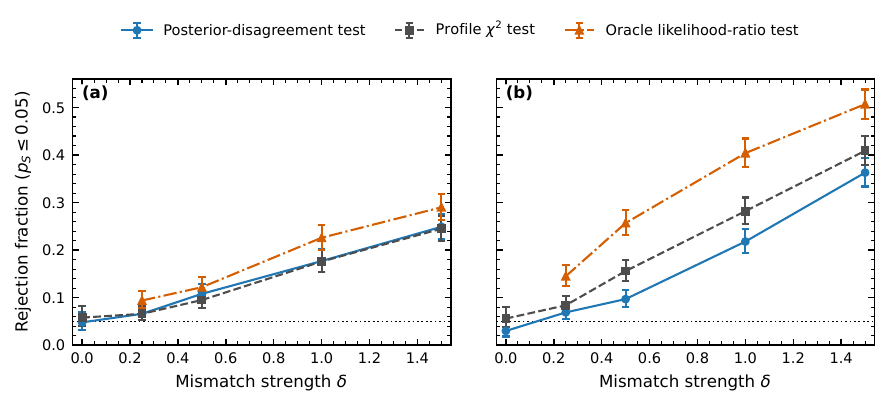}
  \caption{Detection power as a function of mismatch strength $\delta$ for two controlled model-mismatch families. (a) Affine displacement. (b) Nonlinear deformation. For each nonzero $\delta$, rejection fractions are estimated from 1000 independent simulated observations. Uncertainties are 95\% Wilson intervals. The dotted horizontal line marks the nominal false-positive rate of 0.05.}
  \label{fig:toy_global_power}
\end{figure}

The posterior responses in Figure~\ref{fig:toy_geometry} help explain the power curves in Figure~\ref{fig:toy_global_power}. Here, $p_S$ denotes the $p$-value of test $S$, with $p_S=p_{\rm global}$ for the posterior-disagreement test. The rejection fraction is the proportion of simulated observations satisfying $p_S\leq0.05$. At $\delta=1.5$, the posterior-disagreement test rejects 24.9\% of affine-shift observations and 36.3\% of nonlinear-deformation observations. The corresponding fractions are 24.5\% and 40.9\% for the profile-$\chi^2$ test, and 29.0\% and 50.7\% for the oracle likelihood-ratio test, which uses the known injected alternative. All three tests are more sensitive to the nonlinear deformation in these examples. The posterior-disagreement statistic reproduces this distinction without using the alternative model in its construction. The rejection fractions for individual estimator distance combinations at $\delta=1.5$ are shown in Appendix~\ref{app:illustrative_details}. 

\subsection{Validation with simulated CBS data}
\label{sec:simulated_cbs}

We next evaluate the posterior-disagreement test using simulated CBS data. We first check its rejection rate under CPL and then examine its sensitivity to alternative dark-energy histories.

The held-out check rejects 24 of 400 CPL mocks, a fraction of 0.060 with 95\% Wilson interval \([0.0406,0.0877]\). The nominal rejection level of 0.05 lies within this interval, so the observed rejection fraction is consistent with the expected false-positive rate under the CPL reference model. We evaluate the CPL-trained estimator pair against the rapid-transition equation of state introduced by Corasaniti and Copeland~\cite{Corasaniti:2002vg},
\begin{align}
 w_{\rm tr}(a)&=w_0+(w_m-w_0)F(a), \nonumber\\
 F(a)&=\frac{1+\exp(a_c/\Delta)}{1+\exp[-(a-a_c)/\Delta]}
 \frac{1-\exp[-(a-1)/\Delta]}{1-\exp(1/\Delta)}.
 \label{eq:rapid_transition_power_model}
\end{align}
We fix \(w_0=-1\), \(a_c=0.5\), and \(\Delta=0.05\). The remaining cosmological parameters are fixed at \(\omega_b=0.02237\), \(\omega_c=0.1200\), \(H_0=67.4\,{\rm km\,s^{-1}\,Mpc^{-1}}\), \(\tau=0.054\), \(\ln(10^{10}A_s)=3.04452\), and \(n_s=0.965\). A coarse scan over \(-1\leq w_m\leq0\) identifies the response near \(w_m=0\). We resolve this region with a 21-point grid over \(-0.20\leq w_m\leq0\), spaced by 0.01. At each point, the fixed estimators and the CBS calibration ensembles evaluate 200 independent noise realizations, yielding the conditional detection power for this non-CPL departure.

\begin{figure}[H]
  \centering
  \includegraphics[width=0.94\textwidth]{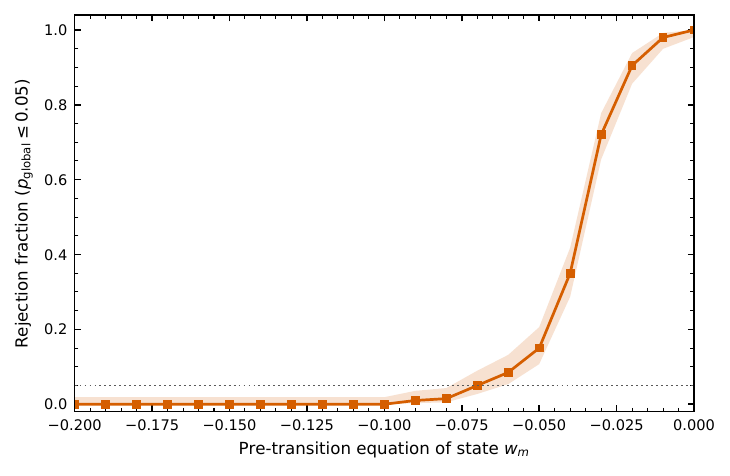}
  \caption{Conditional rejection fraction of the CPL-trained posterior-disagreement test for the CBS data model along the rapid-transition sequence in equation~\eqref{eq:rapid_transition_power_model}, with $w_0=-1$, $a_c=0.5$, and $\Delta=0.05$. Each point uses 200 independent noise realizations at fixed cosmological parameters. The shaded band shows the 95\% Wilson interval, and the dotted line marks the nominal 0.05 rejection level.}
  \label{fig:cpl_trained_noncpl_power}
\end{figure}

Figure~\ref{fig:cpl_trained_noncpl_power} shows the rejection fraction along the tested rapid-transition sequence. The Monte Carlo estimate is first nonzero at $w_m=-0.09$ and reaches 0.150 at $w_m=-0.05$. At $w_m=0$, all 200 noise realizations are rejected, giving a rejection fraction of 1.000 with a 95\% Wilson interval of $[0.981,1.000]$. These results demonstrate a clear response along the tested rapid-transition sequence. Since $a_c$ and $\Delta$ are fixed, varying $w_m$ changes the transition amplitude and the early-time dark-energy evolution, rather than the transition width.

We also consider the smooth Jassal-Bagla-Padmanabhan (JBP) and Barboza-Alcaniz (BA) parameterizations~\cite{Jassal:2005qc,Barboza:2008rh}, defined by $w_{\rm JBP}(z)=w_0+w_a z/(1+z)^2$ and $w_{\rm BA}(z)=w_0+w_a z(1+z)/(1+z^2)$, respectively. At $(w_0,w_a)=(-0.8,-1.0)$, the rejection fraction is 0.015 for each history, estimated from 200 noise realizations. These particular alternatives rarely produce sufficiently large NPE-FMPE disagreement to trigger the test. The contrast with the rapid-transition results shows that a departure from CPL is detectable by this method only when it produces sufficiently unusual disagreement between the two posterior estimators.

\subsection{Application to observed CBS data}
\label{sec:observed_data}

\begin{figure}[H]
  \centering
  \includegraphics[width=0.88\textwidth]{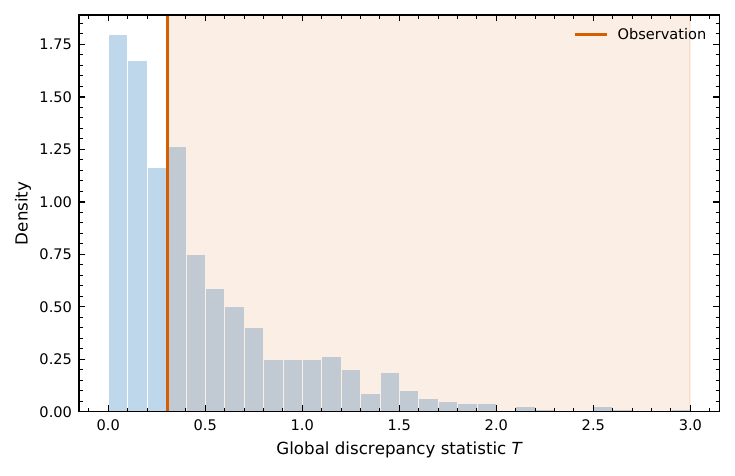}
  \caption{Empirical CPL null distribution of the global disagreement statistic \(T\) for the CBS data. The vertical line marks the observation, and the shaded region marks its upper tail.}
  \label{fig:cpl_global_null}
\end{figure}

Finally, we apply the same CPL-trained estimator pair to the observed CBS vector and assess whether its NPE-FMPE posterior disagreement is unusually large relative to the CPL prior-predictive reference distribution. We obtain $p_{\rm global}=0.5243$. Figure~\ref{fig:cpl_global_null} shows the calibrated reference distribution of the combined disagreement statistic. The observed statistic lies close to the median of the CPL reference distribution, indicating no unusually large NPE-FMPE disagreement for the observed CBS vector.

\begin{table}[t]
\centering
\small
\caption{Marginalized constraints on selected cosmological parameters from the observed-data analyses. Entries give posterior medians and 68\% equal-tailed credible intervals.}
\label{tab:observed_cpl_constraints}
\begin{tabular}{lccc}
\toprule
Parameter & \texttt{Cobaya} & NPE & FMPE \\
\midrule
$\Omega_m$
& $0.315^{+0.008}_{-0.008}$
& $0.315^{+0.008}_{-0.009}$
& $0.311^{+0.008}_{-0.009}$ \\

$H_0\,[{\rm km\,s^{-1}\,Mpc^{-1}}]$
& $66.800^{+0.824}_{-0.811}$
& $66.652^{+0.909}_{-0.777}$
& $67.179^{+0.897}_{-0.754}$ \\

$w_0$
& $-0.790^{+0.081}_{-0.079}$
& $-0.822^{+0.091}_{-0.089}$
& $-0.872^{+0.102}_{-0.092}$ \\

$w_a$
& $-0.639^{+0.272}_{-0.293}$
& $-0.474^{+0.292}_{-0.315}$
& $-0.450^{+0.337}_{-0.330}$ \\
\bottomrule
\end{tabular}
\end{table}

\begin{figure}[H]
  \centering
  \includegraphics[width=0.84\textwidth]{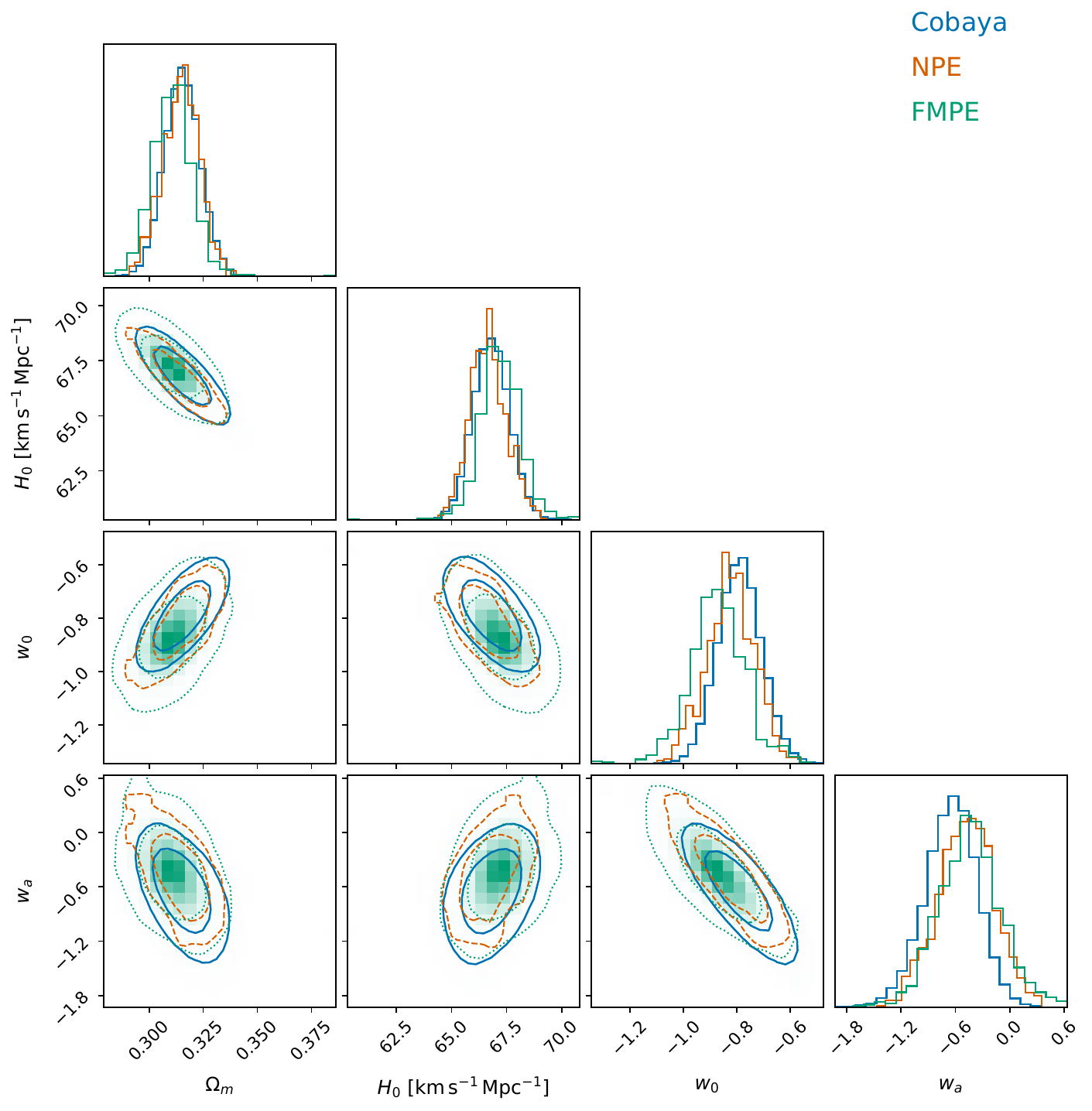}
  \caption{Constraints on cosmological parameters in the CPL model from the CBS data. The blue, orange, and green contours show the results obtained using MCMC (\texttt{Cobaya}), NPE, and FMPE, respectively. The inner and outer contours enclose 68\% and 95\% of the displayed posterior mass, respectively.}
  \label{fig:observed_cpl_corner}
\end{figure}

Figure~\ref{fig:observed_cpl_corner} compares the posteriors of NPE and FMPE inferred from the compressed CBS vector with an independent MCMC result obtained using \texttt{Cobaya}. Table~\ref{tab:observed_cpl_constraints} summarizes the corresponding marginalized constraints on \(\Omega_m\), \(H_0\), \(w_0\), and \(w_a\). All three analyses use the same parameter priors, compressed CBS data vector, and fixed covariance matrix. The MCMC result serves as an external reference and does not enter the calibration of the posterior-disagreement statistic.

The three analyses yield broadly similar central values and overlapping 68\% marginal credible intervals for \(\Omega_m\), \(H_0\), \(w_0\), and \(w_a\). The posterior medians from all three analyses correspond to quintom behavior. This common trend indicates that the qualitative dark-energy evolution inferred from the compressed CBS data is consistent across the neural and MCMC analyses. Once the trained estimators and calibration results are loaded, the online check takes a median of $0.53\,{\rm s}$, allowing efficient repeated evaluations on simulated CBS data.

\section{Conclusion}
\label{sec:conclusion}

In this work, we have developed a simulation-calibrated model check to assess the consistency of the CPL dark-energy model with cosmological observations, using the disagreement between NPE and FMPE posteriors. Two independent CPL mock ensembles establish the expected posterior differences and calibrate their combination into a global $p$-value, without requiring a specified alternative cosmology. Applied to the compressed Planck+DESI+DESY5 data vector, the test gives $p_{\rm global}=0.524$. The CPL model is therefore consistent with the CBS data according to this test.

The controlled experiments connect detection sensitivity to the way model departures affect the estimated posteriors. A shift shared by the estimators produces less disagreement than changes that affect their posterior locations and shapes differently. In the cosmological tests, the rejection fraction reaches 1.00 for the strongest rapid-transition injection, compared with 0.015 for each of the two selected smooth alternatives. These results show that the test identifies departures through their differential effects on NPE and FMPE, with sensitivity varying across the histories and parameter settings examined.

The online model check takes approximately $0.5\,{\rm s}$ once the trained estimators and calibration results are loaded. Most of the computational cost is incurred offline, during simulation, training, and calibration. This enables a rapid, statistically calibrated assessment of posterior disagreement in the observed CBS data.

The different responses to the tested dark-energy histories indicate room to improve the sensitivity of the test. Future work could explore additional SBI methods and posterior-distance measures to detect a wider range of model departures while maintaining statistical calibration. This work provides a way to test for anomalous posterior disagreement under a dark-energy model without specifying an alternative cosmology. Such tests are particularly important for understanding the physical nature of dark energy in the era of precision cosmology.

\section*{Data and code availability}
The data underlying the figures will be made publicly available upon acceptance. The analysis code is available from the corresponding author upon reasonable request.

\section*{Acknowledgments}
This work was supported by the National SKA Program of China (Grants Nos. 2022SKA0110200 and 2022SKA0110203), the National Natural Science Foundation of China (Grants Nos. 12533001, 12575049, 12473001, 12405076, and 12547104), the China Manned Space Program (Grant No. CMS-CSST-2025-A02), and the 111 Project (Grant No. B16009).

\appendix

\section{Compression of the cosmological data vector}
\label{app:implementation}

The cosmological analysis uses a 90-dimensional CBS data vector constructed from CMB, BAO, and SN measurements. This appendix describes the fixed data transformations and covariance treatment used to construct this vector for both simulations and observations.

The CMB spectra are converted to \(D_\ell=\ell(\ell+1)C_\ell/(2\pi)\), and simulations and observations use identical multipole bins. The Planck covariance is evaluated from the released best-fit theory spectrum and \(f_{\rm sky}=0.57\), then projected to unweighted TT, TE, and EE band-power means. The Gaussian approximation is applied throughout \(30\leq\ell\leq900\).

For the SN compression, let $A\in\mathbb{R}^{24\times N_{\rm SN}}$ be the binning matrix, whose nonzero entries are proportional to the inverse statistical variances of the supernovae within each redshift bin and are normalized to sum to unity in each row. Let $C_{\rm SN}$ denote the full statistical-plus-systematic covariance matrix, and define its compressed Cholesky factor $L$ through $LL^T=AC_{\rm SN}A^T$.  An orthonormal basis \(Q_\perp\in\mathbb{R}^{24\times23}\), chosen orthogonal to \(L^{-1}{\bf 1}_{24}\), gives
\begin{equation}
 B=Q_\perp^T L^{-1}A,
 \qquad
 x_{\rm SN}=B\mu_{\rm SN},
 \qquad
 B C_{\rm SN} B^T=I_{23}.
 \label{eq:sn_compression}
\end{equation}
The same fixed transform is applied to simulations and observations.

\section{Finite-sample calibration details}
\label{app:finite_mock}

The global $p$-value is calibrated using the independent second-stage mock ensemble ${\cal O}$. This appendix explains why the resulting empirical rank provides a valid finite-sample test under the CPL prior-predictive reference model, including the treatment of ties and posterior-sampling randomness.

The calibration argument uses the following condition under the CPL reference model:
\begin{equation}
\mathcal{H}_{\rm CPL}:\quad
\bm{x}_{\rm obs},\bm{x}_1,\ldots,\bm{x}_N
\ \text{are exchangeable draws from }p_{\rm CPL}(\bm{x}).
\label{eq:null}
\end{equation}

The marginal rank in equation~\eqref{eq:ps} has resolution $1/(|{\cal R}|+1)$, while the global rank in equation~\eqref{eq:global_p} has resolution $1/(|{\cal O}|+1)$. The observation is included in the rank through the added one in the numerator and denominator, as required for a finite Monte Carlo test. Counting ties with \(\geq\) can only increase the upper-tail rank and is therefore conservative.

The same argument applies after selecting the smallest of the three marginal ranks. Let \(\bm{\xi}\) collect the datum-specific auxiliary random variables used for posterior sampling and, in the controlled nonlinear example, for the SW projection directions. Conditional on the reference set \({\cal R}\), including the auxiliary draws used to construct it, and on any random quantities shared across evaluations, the pairs \((\bm{x}_{\rm obs},\bm{\xi}_{\rm obs})\) and \((\bm{x}_j,\bm{\xi}_j)\) are exchangeable under equation~\eqref{eq:null}. Applying the same map \((\bm{x},\bm{\xi})\mapsto T(\bm{x},\bm{\xi})\) preserves exchangeability. In the absence of ties, the observed statistic is equally likely to occupy any of the \(|{\cal O}|+1\) ranks. With ties, the greater-than-or-equal rule in equation~\eqref{eq:global_p} makes the rank super-uniform and hence conservative. Independent validation mocks assess the rejection rate of the calibrated test and do not enter the calculation of the observed $p_{\rm global}$.

\section{Posterior-coverage diagnostics}
\label{app:posterior_diagnostics}

The posterior-disagreement test measures differences between estimators but does not by itself establish whether their individual posterior approximations have the expected coverage. We examine this separate property using HPD-rank diagnostics on independent simulations.

For the cosmological estimator pair and each independent simulation \((\bm{\theta}_i,\bm{x}_i)\), the HPD rank is
\begin{equation}
 r_{a,i}=\frac{1}{n_{\rm MC}}\sum_{j=1}^{n_{\rm MC}}
 \mathbf{1}\!\left[\log q_a(\bm{\theta}_{ij}|\bm{x}_i)
 \geq\log q_a(\bm{\theta}_i|\bm{x}_i)\right].
 \label{eq:hpd_rank}
\end{equation}
Uniform ranks imply model-averaged coverage~\cite{Talts:2018zdk,Hermans:2021rqv}. We quantify departures from uniformity with
\begin{equation}
 p_{\rm EC}^{(a)}=p_{\rm KS}\!\left(
 \{r_{a,i}\}_{i=1}^{n_{\rm test}},U[0,1]\right),
 \label{eq:coverage_ks}
\end{equation}
where \(p_{\rm KS}\) is the one-sample Kolmogorov-Smirnov (KS) test $p$-value, computed with \texttt{SciPy}~\cite{Virtanen:2019joe}. We use $p_{\rm EC}\leq0.05$ as the pre-specified threshold for identifying a statistically significant departure from uniform HPD ranks.

For the HPD rank diagnostics, the FMPE pointwise log density is evaluated by the continuous-normalizing-flow \texttt{log\_prob} implementation.

For the cosmological estimators, we evaluate HPD ranks on a separate set of 200 CPL simulations, using $n_{\rm MC}=4096$ posterior samples per simulation. For both NPE and FMPE, the KS test gives $p_{\rm EC}>0.05$.

\section{Controlled-validation implementation and supplementary diagnostics}
\label{app:illustrative_details}

This appendix provides the simulator definitions and inference settings for the controlled experiments in section~\ref{sec:illustrative_examples}. It also describes the reference tests and presents supplementary results on null calibration and the rejection fractions of individual estimator-distance combinations.

The one-dimensional example draws \(\theta\sim{\cal U}[-3,3]\) and uses
\begin{equation}
 f(\theta)=\theta,
 \qquad
 g_\delta(\theta)=\theta+\delta,
 \qquad
 x=g_\delta(\theta)+\epsilon,\qquad \epsilon\sim{\cal N}(0,0.30^2).
 \label{eq:toy_affine}
\end{equation}
The training model is \(f=g_0\). Increasing \(\delta\) primarily translates the inferred posteriors while leaving their geometry nearly unchanged.

The two-dimensional example draws independent \(\theta_0,\theta_1\sim{\cal U}[-2.5,2.5]\) and defines the training simulator
\begin{align}
 f_0(\bm{\theta})&=\theta_0+0.25\theta_1^2+0.18\sin(2\theta_0),\nonumber\\
 f_1(\bm{\theta})&=\theta_1+0.35\sin(1.5\theta_0)+0.10\theta_0\theta_1.
 \label{eq:toy_nonlinear_f}
\end{align}
The alternative is \(k_\delta(\bm{\theta})=\bm{f}(\bm{\theta})+\delta\bm{w}(\bm{\theta})\), with
\begin{align}
 w_0(\bm{\theta})&=0.75\sin(2.1\theta_1)+0.28\theta_0\theta_1,\nonumber\\
 w_1(\bm{\theta})&=0.65\cos(1.7\theta_0)-0.22(\theta_1^2-2).
 \label{eq:toy_nonlinear_w}
\end{align}
Data are drawn as \(\bm{x}=k_\delta(\bm{\theta})+\bm{\epsilon}\), with independent Gaussian noise of standard deviation 0.12 in each output coordinate. We use \(\delta\in\{0,0.25,0.5,1.0,1.5\}\). The affine and nonlinear families provide two controlled forms of increasing model mismatch whose posterior responses can be compared directly.

NPE, FMPE, and neural ratio estimation are trained only on the corresponding \(f\) simulator. Their three estimator pairs are combined with the SW, RBF-MMD, and mean-covariance distances, giving nine pre-specified scores,
\begin{align}
 \mathcal{P}&=\{({\rm NPE},{\rm FMPE}),({\rm NPE},{\rm NRE}),
                   ({\rm FMPE},{\rm NRE})\},\nonumber\\
 \mathcal{M}&=\{{\rm SW},{\rm MMD},\mu C\},
 \qquad \mathcal{H}=\mathcal{P}\times\mathcal{M}.
 \label{eq:toy_score_family}
\end{align}
For each \((a,b,m)\in\mathcal{H}\), 5000 simulations in \({\cal R}\) define the marginal rank \(p_{abm}\) using equation~\eqref{eq:ps}. The combined toy statistic is
\begin{equation}
 T_{\rm toy}(\bm{x})=-\log_{10}
 \min_{(a,b,m)\in\mathcal{H}}p_{abm}(\bm{x};{\cal R}).
 \label{eq:toy_global_statistic}
\end{equation}
A 500-mock set \({\cal O}\) calibrates this nine-way minimum with equation~\eqref{eq:global_p}, and a disjoint 500-mock set \({\cal V}\) provides the held-out null check.

Table~\ref{tab:toy_training_settings} summarizes the estimator settings, followed by the implementation, null-calibration, and estimator-distance diagnostics.

\begin{table}[H]
  \centering
  \small
  \caption{Training and sampling settings for the controlled-example estimators.}
  \label{tab:toy_training_settings}
  \begin{tabular}{lcc}
    \toprule
    Setting & Affine & Nonlinear \\
    \midrule
    Training / validation simulations & 200,000 / 50,000 & 200,000 / 50,000 \\
    Hidden layers $\times$ width & $4\times192$ & $4\times192$ \\
    Activation & ReLU & ReLU \\
    Optimizer & AdamW & AdamW \\
    Initial learning rate & $1.5\times10^{-3}$ & $1.5\times10^{-3}$ \\
    Weight decay & $10^{-4}$ & $10^{-4}$ \\
    Batch size & 4096 & 4096 \\
    Maximum epochs & 80 & 80 \\
    Selected epochs (NPE/FMPE/NRE) & 34/43/25 & 24/32/20 \\
    NPE autoregressive transforms & 6 & 6 \\
    Posterior samples & 4096 & 4096 \\
    \bottomrule
  \end{tabular}
\end{table}

Parameters and data are standardized with the training-set means and standard deviations. Training stops after 15 consecutive non-improving epochs once at least 40 epochs have been completed. For the controlled examples, NPE and FMPE operate in unconstrained standardized coordinates, whereas NRE is evaluated on the finite prior grid.

The posterior distances are evaluated in physical parameter coordinates. In the nonlinear example, the SW calculation uses \(L=256\) directions drawn from the fixed run-level pseudorandom stream for each estimator-pair and datum. The RBF bandwidth is the reciprocal median nonzero squared separation of the pooled posterior samples for each estimator pair and datum. MMD uses 1024 samples per estimator. 

We evaluate the NRE likelihood-to-evidence ratio on a uniform prior grid and resample points with probabilities proportional to it. The one-dimensional example uses 20,000 grid points. The two-dimensional example uses a \(241\times241\) grid. Each NRE posterior contains 4096 resampled points. Using the HPD-rank diagnostic defined in Appendix~\ref{app:posterior_diagnostics}, the six estimator-example combinations give \(p\)-values between 0.400 and 0.845, with no statistically significant departure from nominal model-averaged coverage. For a test \(S\), let \(p_S(\bm{x})\) denote its calibrated upper-tail \(p\)-value. For a perturbed data distribution \(p_\delta\), the detection probability at level \(\alpha\) is
\begin{equation}
 {\cal P}_S(\delta;\alpha)=
 P_{\bm{x}\sim p_\delta(\bm{x})}
 \!\left[p_S(\bm{x})\leq\alpha\right].
 \label{eq:power_curve}
\end{equation}

For the posterior-disagreement test, \(p_S=p_{\rm global}\).

The profile reference statistic is
\begin{equation}
 \chi^2_{\rm prof}(\bm{x})=
 \min_{\bm{\theta}\in\Theta}
 \frac{\|\bm{x}-\bm{f}(\bm{\theta})\|_2^2}{\sigma^2}.
 \label{eq:toy_profile_chi2}
\end{equation}
Its upper-tail rank is calibrated with 5000 simulations from \(\bm{f}\), and its zero-mismatch rejection fraction is evaluated on 500 independent null observations. For the oracle reference, define
\begin{equation}
 h_\delta(\bm{\theta})=
 \begin{cases}
  g_\delta(\bm{\theta}), & \text{affine example},\\
  k_\delta(\bm{\theta}), & \text{nonlinear example}.
 \end{cases}
 \label{eq:toy_alternative_symbol}
\end{equation}
The oracle likelihood-ratio statistic is
\begin{equation}
 \Lambda_\delta(\bm{x})=
 \log p_{h_\delta}(\bm{x})-\log p_f(\bm{x}),
 \qquad
 p_h(\bm{x})=\int p_h(\bm{x}|\bm{\theta})\pi(\bm{\theta})\,d\bm{\theta}.
 \label{eq:toy_oracle_lr}
\end{equation}
We approximate the minimization and marginalization on the same uniform grids used for NRE sampling and calibrate the upper tail of \(\Lambda_\delta\) with the null simulations. Because the injected alternative and \(\delta\) are known, \(\Lambda_\delta\) provides an oracle reference for the achievable detection sensitivity.

\begin{figure}[H]
  \centering
  \includegraphics[width=0.8\textwidth]{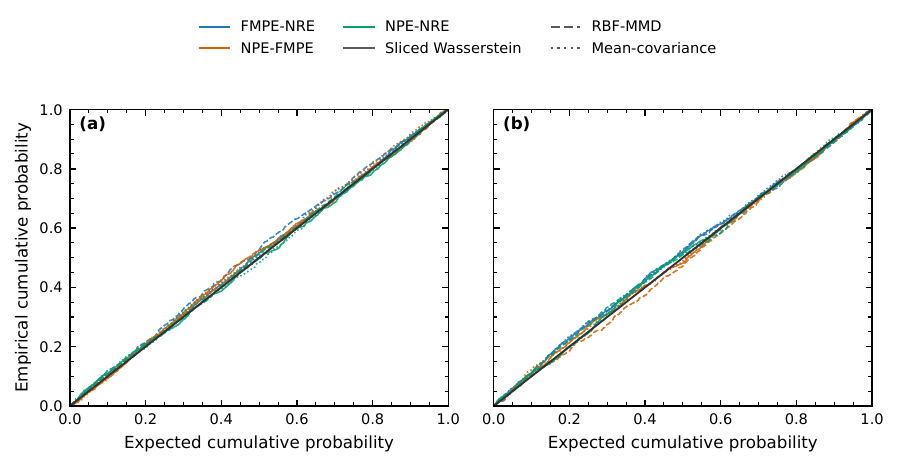}
  \caption{P-P plots of the individual $p$-values $p_{abm}$ under zero mismatch. (a) Affine displacement. (b) Nonlinear deformation. The black diagonal shows the uniform reference.}
  \label{fig:toy_null}
\end{figure}

The P-P plots under zero mismatch use 1000 zero-mismatch observations for each example. Figure~\ref{fig:toy_null} shows their empirical cumulative distribution functions.

\begin{figure}[H]
  \centering
  \includegraphics[width=0.8\textwidth]{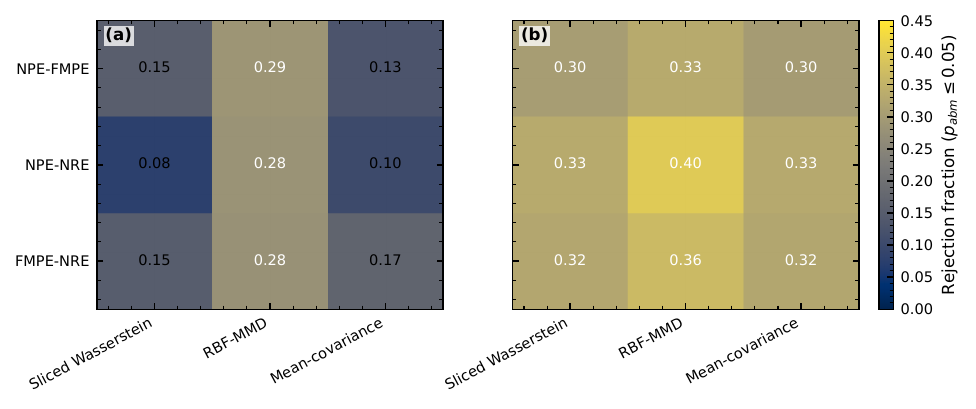}
  \caption{Estimator-distance rejection-fraction matrices at $\delta=1.5$. (a) Affine displacement. (b) Nonlinear deformation. Rows denote estimator pairs and columns denote posterior distances.}
  \label{fig:toy_robustness}
\end{figure}

The individual ranks follow the expected uniform distribution in both examples. For the estimator-distance comparison, each cell is the fraction of 1000 observations with \(p_{abm}\leq0.05\) at \(\delta=1.5\). Figure~\ref{fig:toy_robustness} shows this matrix.

\bibliographystyle{JHEP}
\bibliography{main}

\end{document}